\documentstyle[fleqn]{tp}

\input psfig

\begin{document}

\twocolumn[
\title{Cold Flows and Large Scale Tides}
\author{Rien van de Weygaert$^1$ and Yehuda Hoffman$^{1,2}$\\
{\it $^1$Kapteyn Astronomical Institute, P.O. Box 800, 9700 AV Groningen, the Netherlands}\\
{\it $^2$Racah Institute of Physics, The Hebrew University, Jerusalem 91904, Israel}}
\vspace*{16pt}   

ABSTRACT.\
We propose a different view of the dilemma concerning the 
coldness of the local cosmic flow and its repercussions for the global 
Universe. We stress the 
fact that our cosmic neighbourhood embodies a region of rather particular 
circumstances, whose dynamics and kinematics are substantially influenced 
by its location in between the Great Attractor region and the Pisces-Perseus 
chain. On the basis of constrained simulations of our cosmic neighbourhood 
we indicate through the cosmic Mach number that we live in an extraordinarily 
cold niche of the Universe.
\endabstract]

\markboth{van de Weygaert \& Hoffman}{Cold Flows and Large Scale Tides}

\small

\begin{figure*}
\centering\mbox{\hspace{-3.5cm}\psfig{figure=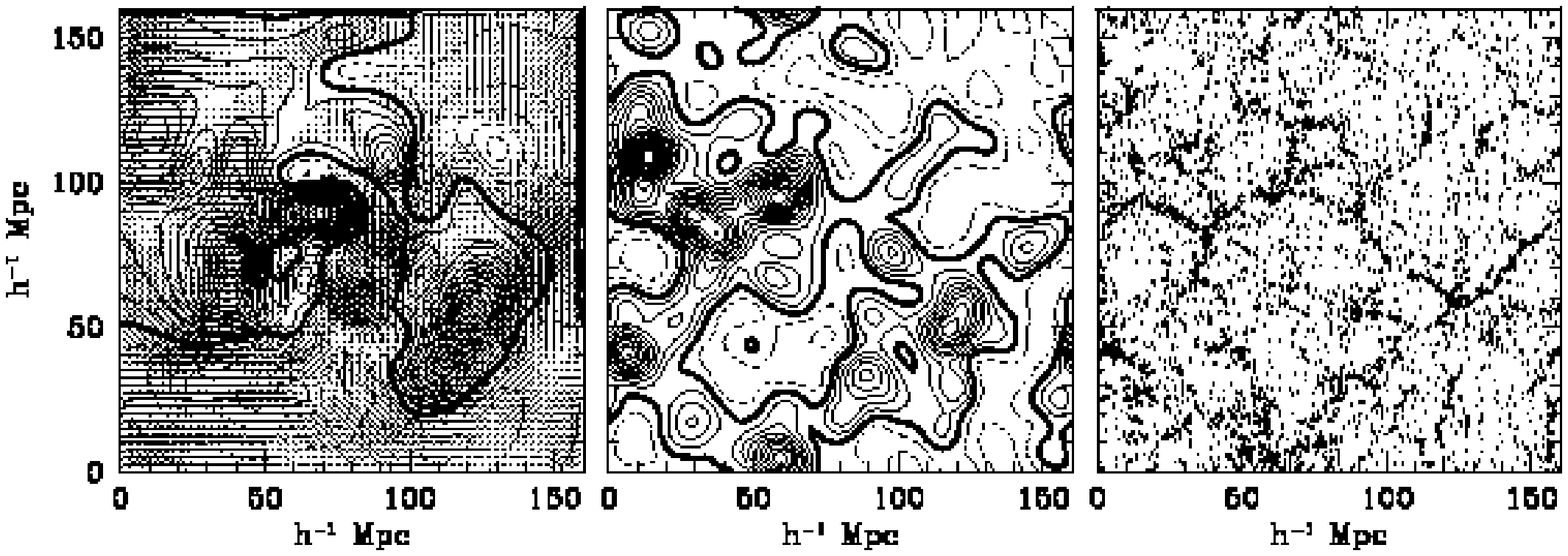,width=18.cm,angle=270.0}}
\caption[]{\hfill\\
\addtolength{\baselineskip}{-5pt}
\vbox{Left: The density distribution in a constrained reconstruction 
of our Local Universe. Based on the Mark III catalogue peculiar motions, 
smoothed on a Gaussian scale of $R_f=5h^{-1} \hbox{Mpc}$,  
the Wiener filter reconstruction yields the density field in the top left 
frame, represented by a density contour plot in the central $x-y$ plane. 
The solid contours correspond to $\delta > 0.0$, the dashed ones to 
low-density regions with $\delta < 0.0$. The indicated density levels range 
between -0.6 and 1.3, with the linear increment 
between contour levels amounting to $\Delta \delta = 0.1$. The heavy 
solid line is the $\delta=0.0$ contour. 
We, the Local Group, are at the centre of the box, clearly recognizable 
are the mass concentrations around the Perseus-Pisces complex (lower 
lefthand side) and the Great Attractor region (upper righthand corner). 
Superimposed on the Wiener filtered density field contours are 
the bars representing the compressional component of the tidal shear within 
the $x-y$ plane (see van de Weygaert \& Hoffman 1998). Notice the 
strength of the tidal field in the underdense 
region near our own cosmic location. Centre: the 
constrained linear density field realization after including a 
constrained contribution of small-scale waves according to the standard 
CDM scenario, Gaussian smoothed on a scale of $R_f=5h^{-1} \hbox{Mpc}$. 
The density contours are characterized in the same way as 
the ones in the top frame, with the linear increment of the contour 
levels amounting to $\Delta \delta = 0.2$, ranging in value from 
$\delta = -1.6$ to $\delta =3.0$. Right:  outcome of the 
nonlinear evolution 
of the constrained realization of our local Universe, as evolved by means 
of an N-body simulation.}\addtolength{\baselineskip}{5pt}}
\end{figure*}

\section{Cosmic Chills and Universal Truths}
Early assessment of the small-scale random motions of galaxies, estimated 
on the basis of pairwise velocity dispersions, revealed that locally the 
Universe is rather cold. While we participate in a bulk flow of 
approximately $600 \hbox{km/s}$, the random velocities with respect 
to the mean flow are estimated to be in the range of a mere 
$200--300 \hbox{km/s}$ (see e.g. Suto, Cen \& Ostriker 1992)
This low value of the velocity dispersion in combination with the pronounced 
structure displayed by the distribution of galaxies was in fact a strong 
argument for either a low $\Omega$-Universe the or for the latter being a 
biased tracer of the underlying matter 
distribution. The assumption of bias, in particular in the form of the 
oversimplifying linear bias factor $b$, would then imply the matter 
distribution not to have evolved as far as suggested by the pronounced 
nature of the galaxy distribution, and hence would be in agreement with 
the low value of the ``thermal'' motions in the local Universe.  

An interesting elaboration on the assessment of the implications of the 
coldness of the local velocity flow for the properties of the global 
Universe, in particular for the value of the cosmological density 
parameter $\Omega$, was suggested by Ostriker \& Suzo (1990). They pointed 
out that a comparison between the properties of the small-scale 
``dispersion'' velocities, $\sigma({\bf v})$, and the large-scale bulk 
motions, $|{\bf v}|_{bulk}$, would not only 
provide valuable information on the relative amount of large-scale and 
small-scale power in the spectrum of density- and velocity fluctuations, but 
that to do this on the basis of their ratio, the ``cosmic Mach number'' 
${\cal M}$, 
\begin{equation}
{\cal M} \equiv {|{\bf v}|_{bulk} \over \sigma({\bf v})}\,.
\label{eq:1}
\end{equation}
has the additional advantage of providing this information in a way 
that is independent of both the -- not yet definitively settled -- amplitude 
normalization of the power spectrum as well as of a possible -- linear, 
scale-indepent -- bias between galaxies and matter. The only intrinsic 
assumption is that the velocities 
of galaxies do form an unbiased tracer of the underlying velocity field. 
One of the most striking conclusions 
reached on the basis of extensive tests of the discriminatory virtues 
of the Mach number was that the velocity field in our local Universe 
appeared to exclude the viability of the standard CDM model 
(Ostriker \& Suzo 1990).

\section{Local versus Global}
In an attempt to interpret the significance of the coldness of the 
local cosmic flow, we postulate an alternative view. Rather than interpreting 
the coldness of the flow as a property of the global Universe, we hold the 
view that it is our rather uncommon cosmic location that lies at the basis of 
the issue and that we should be careful in drawing conclusions concerning 
the global value of cosmic parameters as well as of the validity of 
structure formation scenarios. 

\begin{figure*}
\centering\mbox{\hspace{-2.5cm}\psfig{figure=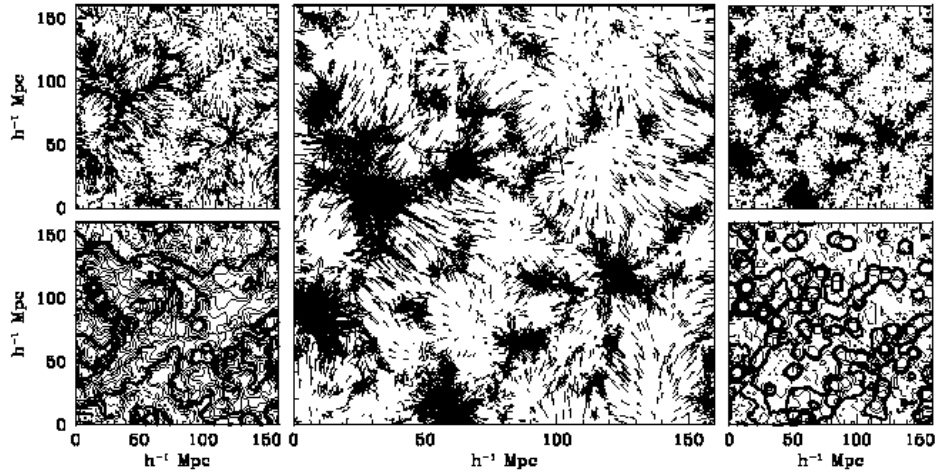,width=15.cm,angle=270.0}}
\caption[]{\hfill\\
\addtolength{\baselineskip}{-5pt}
\vbox{Notice $90^{\circ}$ change in figure orientation. The velocity field 
structure in our local Universe. Central 
frame: full velocity field at particle locations, represented by velocity 
vectors of velocity components in central $x-y$ plane. Left frames 
contain the bulk velocity component of the velocity field at the 
particle locations, with the top frame displaying the vector 
representation, and the bottom frame the contour plot of the amplitude 
of the bulk velocity in the central $x-y$ plane. The righthand frames 
display the same for the small-scale velocity `dispersion' component. 
The bulk velocity is defined as the velocity after filtering out the 
contributions on scales smaller than $5h^{-1} \hbox{Mpc}$, while the 
velocity dispersion is the remaining small-scale residual velocity.}
\addtolength{\baselineskip}{5pt}}
\end{figure*}

Figure 1a contains a reconstruction of the linear density field 
(Gaussian scale $R_f = 5h^{-1}\,\hbox{Mpc}$) in 
our local Universe, in a slice approximately coinciding with the 
Supergalactic Plane, based on the set of measured peculiar velocities 
of galaxies in the Mark III catalogue (Willick et al. 1997). The 
Local Group is located at the center of the box. On the upper lefthand 
side we can discern a huge positive density complex, the Great Attractor 
region, while on the other side, lower righthand corner, we observe the 
presence of another massive density enhancement, corresponding to the 
Perseus-Pisces supercluster region. Moreover, in perpendicular directions
we have vast regions of lower than average density. Hence, seemingly  
we find ourselves located right near the centre of a configuration strongly 
reminiscent of a canonical quadrupolar mass distribution. The direct 
dynamical implication of this is that we are located near the saddle point 
of a strong field of tidal shear. In fact, when turning to Figure 1b we 
see the compressional component of the tidal field corresponding to the 
density distribution in the local Universe (Van de Weygaert \& Hoffman 
1998), superposed on the isodensity contours in the same slice. The tidal 
field, illustrated by bars whose size and direction are proportional to 
the strength of the compression along the indicated direction is, evidently,
very strong within the realm of the two huge matter concentrations where 
the density reaches high values. In addition, however, we also see that 
the tidal shear is indeed very strong at our own location, precisely 
due to the fact that we are hanging in between the Great Attractor 
and the Perseus-Pisces chain. In fact, through the anisotropic nature 
of the induced tidal forces we can recognize a situation that was for 
instance already recognized within the context of the Cosmic Web 
picture (see Bond, Kofman \& Pogosyan 1997, also see van de 
Weygaert \& Bertschinger, 1996, fig. 4), the collapse and formation 
of filamentary and tenuous wall-like features in regions bordered 
by massive matter clumps like clusters of galaxies. This appears to 
be the generic outcome of the evolutionary path of regions of moderately 
low overdensities immersed in an external tidal field whose strength  
is of the same order of magnitude as the selfgravity of the region. In 
such situations we may expect the contraction itself to be 
accelerated with respect to the situation of the same initial region 
having been spatially isolated. In fact, the shearing nature of the 
forces will accelerate the contraction along the shortest axis of the 
region that will experience an accelerated contraction, while 
the longest axis will experience a slow down of its contraction 
rate. This in turn will very likely imply a slower mixing, ``thermal'' 
settling and virialization of the matter involved in the contraction process. 

It is the preceding sketch of events that prompts us to expect our 
local corner of the Universe, because of its special location in 
between the Great Attractor and the Perseus-Pisces complex, to 
have a substantially lower velocity dispersion than we may expect 
to encounter in an average patch of our Universe.

\section{Wiener filtering the Mark III Local Cosmos} 
In an attempt to address the implications of the coldness of the 
local cosmic flow, we investigated the dynamical and kinematical 
evolution of cosmic regions resembling as closely as possible our 
own local Universe. 

First issue in this approach is to set a cosmic environment 
resembling our local Universe. This is achieved by invoking relevant 
observationally determined properties of the local cosmos. As we are 
specifically interested in dynamical issues, we base ourselves on 
a local cosmic primordial linear density field that represents 
an optimally significant reconstruction of the prevailing matter 
distribution in the early Universe. We achieve this by applying a 
Wiener filter algorithm to the sample of measured peculiar 
velocities of galaxies in the Mark III catalogue (Willick et al. 
1997, and Zaroubi et al. 1995 for technical aspects). 
Such a reconstruction restricts itself to regions that arguably are 
still within the linear regime, and whose statistical properties are still 
Gaussian. We discard further observational constraints on the small-scale 
clumpiness and motions in the local Universe. Instead, we generate 
and superpose several realizations of small-scale density and 
velocity fluctuations according to a specific power spectrum 
of fluctuations, global cosmological background specified by 
$H_0$ and $\Omega_0$, and with the small-scale noise being 
appropriately modulated by the large-scale Wiener filter 
reconstructed density field. To this end we invoke the technique 
of constrained random fields (see Hoffman \& Ribak 1991, van 
de Weygaert \& Bertschinger 1996), with the Wiener filtered 
field playing the role of ``mean field''.

\begin{figure*}
\centering\mbox{\hspace{-1.0cm}\psfig{figure=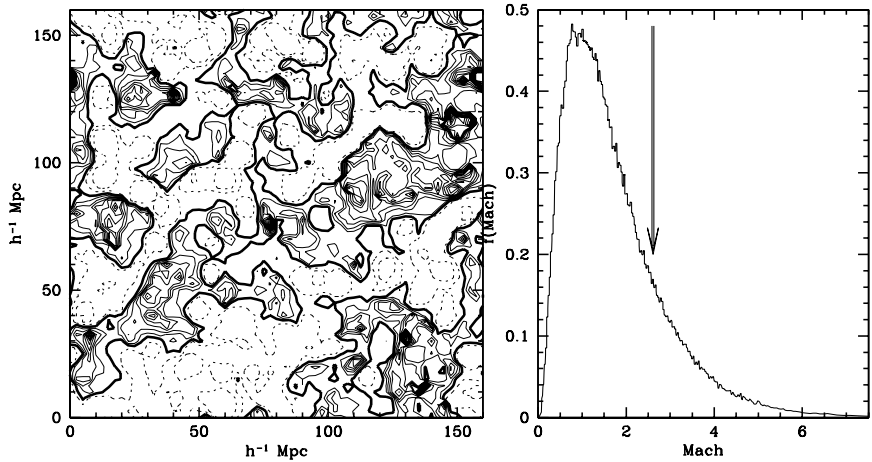,width=12.cm}}
\caption[]{\hfill\\
\addtolength{\baselineskip}{-5pt}
\vbox{Contour plot of the Mach number distribution in the central 
$x-y$ slice, together with the probability distribution function of the 
Mach number (bottom frame). The arrow indicates the location of the 
value of the Mach number at our cosmic location (centre of box), equal 
to $M=2.61$ for the presented Local Universe simulated realization.}
\addtolength{\baselineskip}{5pt}}
\end{figure*}

\section{Small-scale evolution of the Local Cosmos} 
Having generated a full realization of a patch of the Universe 
resembling the primordial density field in our local cosmic neighbourhood, 
we trace its development by means of an P$^3$M N-body simulation.
The outcome of our simulations is reduced and analyzed with the help of a 
`dynamical fields' code. The resulting distribution 
of the particles in a central slice through the simulations box is shown 
in the central frame of Figure 2. Clearly recognizable are massive 
concentrations of matter at the locations where in the real Universe 
we observe the presence of the Great Attractor region (slightly 
``north'' of the ``west'' direction) and the Perseus-Pisces region 
(slightly ``south'' of the ``east''). Interesting is to see how 
vast and extended these regions in fact are, certainly not to be 
identified with well-defined singular objects. The corresponding velocity 
field in the same region of the simulations is displayed by means of a 
decomposition in the large-scale bulk flow ${\bf v}_{bulk}$, top-hat 
filtered in a sphere of radius $R_{TH}=5h^{-1}\hbox{Mpc}$, and the 
small-scale velocity ``dispersion'' ${\bf v}_{\sigma}$ (for technical 
details see Van de Weygaert \& Hoffman 1998). The 
top-lefthand frame and the top-righthand frame contain contour plots 
of the amplitude of these velocity field components, while the lower lefthand 
and lower righthand frame show the corresponding vectorial representation 
of the velocities at the locations of the particles. In particular the 
bulk flow field provides a beautiful impression of the displacement 
of matter towards the emergence of large-scale features like filaments 
and voids. Most interesting though is that also the small-scale dispersion 
field appears to bear the marks of underlying large-scale features: not 
only do we see large ``thermal'' velocities at the sites of cluster 
concentratios, but we can also recognize sizeable small-scale velocities 
near the locations of filaments (see Van de Weygaert \& Hoffman 1998). 

When we compare the contour plots of the bulk motion and the dispersion 
velocities, we can already discern the fact that while we (i.e. the 
centre of the simulation box) are still embedded in a region of high bulk 
flow, evidently incited by the GA and the PP region, we also find ourselves 
in a region of exceptional low velocity dispersion. Evidently a nice 
reflection of the observed ``coldness'' of the local cosmos. We should not 
fail to notice that apparently these small-scale repercussions are the natural 
outcome of the dynamical evolution of a region possessing the large-scale 
features ($R > 5h^{-1} \hbox{Mpc}$) of the local universe. 

Secondly, an inspection of the contour map of the spatial comsic Mach number 
distribution illuminates the significance of the ``coldness'' of the 
local cosmic flow. A comparison with the density map in Figure 1 reveals 
the interesting aspect of a large coherent band of high Mach number values 
running from the lower lefthand side to the upper righthand side of 
the simulation box, avoiding both the Great Attractor region and the 
Pisces-Perseus region, situated approximately in between those two complexes, 
almost along the bisecting plane that they define (see van de Weygaert 
\& Hoffman 1998). Superposed on this large-scale pattern are a plethora 
of small-scale features. For our purpose the most significant of these is 
the fact that we appear to be right near a towering peak of the Mach 
number distribution. In fact, for this N-body realization of our 
Universe we find at the location of the Local Group a bulk velocity of 
$695.5\hbox{km/s}$, a velocity dispersion of $266.5\hbox{km/s}$ and 
a Mach number of $2.61$, corresponding to a percentile level of $82.1\%$. 
That point is stressed further by invoking the statistical distribution 
function of the Mach number in Figure 3b. Clearly we find ourselves 
somewhere in the tail of the distribution, potentially rendering it 
possible that we do live in a high-density Universe even though locally 
it's chilly ... Although the exact numbers differs for various realizations, 
and for instance the spatial patterns in the Mach number distribution 
amy display different features (narrower band, no conspicuous peak), the 
Mach number consistently assumes a value in the range above the $80\%$ 
percentile value.

\end{document}